\documentstyle[12pt]{article}
\begin{document}

\rightline{FTUV/95-23}
\rightline{IFIC/95-23}
\rightline{June 9, 1995}

\begin{center}
{\large{ \bf Group Theoretical Foundations of Fractional Supersymmetry}}
\end{center}

\vspace{1cm}

\begin{center}
{\large {J. A. de Azc\'arraga and A. J. Macfarlane\footnote{
On leave of absence from DAMTP, Cambridge University,
Silver Street, Cambridge CB3 9EW, UK, and from
St. John's College, Cambridge, UK.}}}
\end{center}

\begin{center}
{\small{\it Departamento  de F\'{\i}sica Te\'orica and \\
IFIC (Centro Mixto Universidad de Valencia-CSIC) \\
46100 Burjasot (Valencia), Spain}}
\end{center}

\vspace{3cm}

\begin{abstract}
{\small{\
Fractional supersymmetry denotes a generalisation of  supersymmetry
which may be constructed using a single real generalised Grassmann variable,
$\theta = \bar{\theta}, \, \theta^n = 0$, for arbitrary integer
$n = 2, 3, ...$. An explicit formula is given in the case of general
$n$ for the transformations that leave the  theory invariant, and it is
shown that
these transformations possess interesting group properties.  It is
shown also that the two generalised derivatives that enter the theory
have a geometric interpretation as generators of left and right
transformations of the fractional supersymmetry group. Careful attention
is paid to some technically important issues, including differentiation,
that  arise as a result of the peculiar nature of quantities such as $\theta$.
}}
\end{abstract}

\newpage

\section{Introduction}

Supersymmetry has been a popular and fruitful area of research for at least
twenty years. Study of it in space-time of one dimension, time, has given rise
to the important topic of supersymmetric quantum mechanics (see \cite{w}-
\cite{gk}
 for
reviews). The most primitive version of supersymmetric quantum mechanics is
one that involves use of a single real Grassmann number $\theta$ such that
$$
\theta=\bar{\theta} \quad , \quad \theta^2=0 \quad .
\eqno{(1.1)}
$$
\noindent
As a result the theory possesses a natural ${\cal Z}_2-$grading and a single
generator $Q$ of its supersymmetry transformations which obeys
$Q^2=-\partial_t$. The distinctive features of supersymmetric
theories which possess such a ${\cal Z}_2-$grading can be seen by reference
to various papers \cite{sa}-\cite{mf}.
The term fractional supersymmetry is currently being
applied to a class of generalisations of supersymmetry in one dimension. Our
work on fractional supersymmetry can be presented most straightforwardly by
creating theories with ${\cal Z}_n-$grading by generalisation of theories with
${\cal Z}_2-$grading.
Thus we consider theories involving a single real
(generalised) Grassmann number $\theta$ which obeys
$$
\theta=\bar{\theta} \quad , \quad \theta^n=0 \quad ,
\quad n=2,3,4, \dots \quad ,
\eqno{(1.2)}
$$
\noindent
in which the generator $Q$ of the generalised (`fractional') supersymmetry
transformations that leave  such a theory invariant obeys
$$Q^n=-\partial_t \quad .
\eqno{(1.3)}
$$
\noindent
The last result accounts loosely for the use of the term `fractional' as an
identifier of the theory.

The generalisation from ordinary  to fractional
supersymmetry not only has intrinsic interest but may also be
expected to produce interesting new models in classical and quantum
mechanics. There have been a large number of studies of fractional
supersymmetry in recent years \cite{bf} - {\cite{cm}. Some of these
deal with a complex Grassmann variable $\theta$ such that $\theta^n
= 0$. Others employ $N$ different copies of $\theta$ which obey
(1.1), thus developing $N$-extended
fractional supersymmetry. Fractional supersymmetry is contrasted below with
a distinct class of generalisations of basic, or
${\cal Z}_2$ -graded, supersymmetry, those which possess parasupersymmetry.
There has been a great deal of attention given recently to work
 in this field \cite{rs} - \cite{nde}, and often these papers contain
thinking relevant also to fractional supersymmetry. We belive that the
whole  area promises both activity and progress in the future.

This paper discusses two important aspects of fractional
supersymmetry,  Firstly, we discuss the fact that the fractional
supersymmetry transformations that describe the invariance
properties of the ${\cal Z}_n$-graded theory form a group
$ G_n$. Secondly, we elucidate
certain fundamental technical matters stemming from  unfamiliar features
of the algebra. Two areas  need attention. One
concerns differentiation with respect to $\theta$; the other is the situation
surrounding families of multiplicative rules of the type

$$
\epsilon \theta = q^{-1} \theta \epsilon \quad  ,
\quad q = \exp (2 \pi i/n) \quad ,
\eqno{(1.4)}
$$

\noindent
involving a Grassmann number $\theta$, its associated
 transformation parameter $\epsilon$
 and dynamical variables of Grassmann type. In
the former there are difficulties of principle, which we treat; in the latter
it is a matter of demonstrating a coherent rationale behind the formulation
and the consistency of results like (1.4) within it.

We give explicit formulas for the elements of the group $G_n$ of
transformations that should leave any  ${\cal Z}_n$-graded theory invariant,
and related proofs. Once the status of derivatives with respect to
$\theta$ is established, we turn to the
objects of generalised covariant derivative type that enter (up to now in
an {\it ad hoc} way) into a  theory possessing
fractional symmetry. We show that these have the
interesting geometrical interpretation of being the generators of the
left and right actions of the fractional supersymmetry group $G_n$
(as is the case for ordinary supersymmetry \cite{aa}).

We have introduced into our
discussion a quantity $q = \exp (2 \pi i/ n)$ which obeys

$$
q^n = 1 \quad .
\eqno{(1.5)}
$$

\noindent
To provide some appropriate comment, we recall  that
fractional symmetry  aims at a generalisation of supersymmetry. The
latter when quantized, involves one boson and
one fermion variable, and requires use of a $2\times2$ matrix representation
of the
fermion. We plan a generalisation, (see (4.3), (4.1) for $n = 3$
or (7.2) below), which retains the boson and replaces the fermion
by some more general object, cf. (1.1) and
(1.2). Two classes of variables, which can be
represented by matrices in an $n$-dimensional vector space are known to us.
The parafermions \cite{gr}\cite{ok} are one of these; use of them leads
down a path of interest, but not the one we are able usefully to
follow at the moment, toward parasupersymmetry. We follow the other path. The
$q$-deformed harmonic oscillator \cite{ac}\cite{mac}\cite{bi}
possesses commutation relations in terms of its $a$
and $a^{\dagger}$ variables that make sense, not only for
$q \in  R$, but also $q \in  C$, when (1.5)
applies. In this situation,
$a$ and $a^{\dagger}$ are represented (for each $n$) by $n \times n$ matrices.
Since for $n
= 2$ we get back in this way to a description of fermions, it is clear that
we are talking about generalisations of these. By looking at the $n = 3$ case
and  beyond  one can see that the generalisations are distinct from
parafermions. We are not yet in a position
to push satisfactorily the quantization
of our theory to a point where the implied interpretation is present in
a consistent well-understood way, but we are certainly describing a
plausible scenario for it. We plan to report on this soon.

For reasons of notational simplicity and clarity,
we present first our ideas for the  ${\cal Z}_3$-graded
case, the first non-trivial generalisation of the basic supersymmetry.
This already requires that most of the central issues of the ${\cal Z}_n$ case
be treated seriously.
The paper contains seven sections. Section 2 contains
introductory material for $G_3$ including its group law, and the
reasons behind expressions such as (1.4). Section 3 derives the formula for the
transformations of  $G_n$. Section 4
discusses the problem of defining
derivatives with respect to $\theta$, leading into section 5
which shows how the usual derivatives $Q$ and $D$ enter crucially into the
construction of a Lagrangian theory with $G_3$ invariance. Section 6
establishes $Q$ and $D$ as the generators of the left and right actions of
$G_3$ by introducing a suitable exponentiation of the first
order formulas. Section 7
extends our results for $n = 3$ to the general case and includes
the proof of the exponentiation for general $n$.

\section{Fractional supersymmetry transformations: the  case of $G_3$}

The simplest version of ordinary supersymmetry deals with the transformation

$$
t' = t + \tau + i \epsilon \theta \quad , \quad \theta' = \theta + \epsilon
\quad .
\eqno{(2.1)}
$$

\noindent
This ${\cal Z}_2$-graded
theory contains a  time variable $t$
and a parameter $\tau$ of grade zero, and a real Grassmann number
$\theta$ and parameter $\epsilon$ of grade one. Thus

$$
\theta = \bar{\theta} \quad , \quad \theta^2 = 0
\quad ;  \quad \epsilon = \bar{\epsilon}
\quad , \quad \epsilon^2 = 0 \quad ; \quad \theta \epsilon = - \epsilon
\theta \quad .
\eqno{(2.2)}
$$

\noindent
We consider generalisation to a situation involving a single real Grassmann
variable $\theta$, such that $\theta = \bar{\theta}$, $\theta^n = 0$ ,
$n = 2,3,4 ...$, within a theory that possesses ${\cal Z}_n$-grading; the
case $n = 3$ provides the simplest non-trivial generalisation of
ordinary supersymmetry.
Without loss of generality, we take  $\theta$ to have
grade one in the ${\cal Z}_3$-grading, and  to obey

$$
\theta = \bar{\theta} \quad , \quad \theta^3 = 0 \quad .
\eqno{(2.3)}
$$

\noindent
The ${\cal Z}_3$-generalisation of (2.1) is then given by

$$
t \rightarrow t'  = t + \tau + \xi (\epsilon, \theta) \quad ,  \quad
\theta \rightarrow \theta'  = \theta + \epsilon \quad ,
\eqno{(2.4)}
$$

\noindent
where

$$
\xi (\epsilon, \theta) = q (\epsilon \theta^2 + \epsilon^2 \theta) \quad ,
\eqno{(2.5)}
$$

\noindent
$t$ and $\tau$ are as in (2.1), $\epsilon$ is a real grade one parameter,
such that $\epsilon = \bar{\epsilon} \, , \, \epsilon^3 = 0$,

$$
\epsilon \theta   = q^{-1} \theta \epsilon \quad .
\eqno{(2.6)}
$$

\noindent
and  $q$ is a complex cube-root of unity. For definiteness
we take $q = \exp (2 \pi i /3)$; replacing
$q$ by $q^{-1}$ in (2.5) and (2.6) would modify only slightly the appearance
of the expressions written below, but not their content.
Eq. (2.6) ensures that the two terms of (2.5), in addition to being
of overall grade zero, are real, {\it e.g}.

$$
\overline{q \epsilon \theta^2} = q^{-1} \theta^2 \epsilon = q^{-1} q^2 \epsilon
\theta = q \epsilon \theta^2 \quad .
$$

The fact that (2.4) describes a group $G_3$ of transformations is easy to
check. Applying two transformations  $g = (\tau, \epsilon)$ and
$g' = (\tau', \epsilon')$ to $(t, \theta)$ we find
$ g'' = g'g$ with parameters

$$
\epsilon'' = \epsilon' + \epsilon \quad , \quad
\tau '' = \tau ' + \tau + q (\epsilon' \epsilon^2 + \epsilon'\,^2 \epsilon)
\equiv  \tau' +
\tau + \xi (\epsilon', \epsilon)
\eqno{(2.7)}
$$

\noindent
where, in analogy with (2.6), we have

$$
\epsilon' \epsilon = q^{-1} \epsilon \epsilon' \quad .
\eqno{(2.8)}
$$

\noindent
In fact, we may view (2.4) as the (left) action of the element
$g \in G_3$ on a ${\cal Z}_3$-graded
 physical `manifold' $M$, of  `coordinates' $(t, \theta)$,
given by

$$
g: (t, \theta) \mapsto (t' \theta') \quad , \quad
\theta' = \theta + \epsilon \quad , \quad t' =  t + \tau + \xi
(\epsilon, \theta) \quad ;
\eqno{(2.9)}
$$

\noindent
likewise, we may view (2.7) as describing the left action of $g' = (\tau',
\epsilon')$ on the $G_3$ group itself.

The unit and inverse elements are given by (0,0) and $(- \tau,
- \epsilon)$. The associativity of the group law $g'' (g' g) = (g'' g')g$
 is easily
checked, and, in fact, it follows from two-cocycle condition

$$
\xi (\epsilon'', \epsilon') + \xi (\epsilon''+ \epsilon',
\epsilon) = \xi (\epsilon'', \epsilon'+
\epsilon)+ \xi (\epsilon', \epsilon) \quad ,
\eqno{(2.10)}
$$

\noindent
in which $\epsilon, \epsilon', \epsilon''$ are the grade one
parameters of three transformation performed in succession, and which
holds for the $\xi (\epsilon' , \epsilon) $ given in (2.7)
provided that

$$
\epsilon'' \epsilon' = q^{-1} \epsilon' \epsilon \quad , \quad
\epsilon' \epsilon = q^{-1} \epsilon \epsilon' \quad .
\eqno{(2.11)}
$$
\noindent
As is well known (see, {\it e.g.} \cite{ai}), two-cocycles are associated
with central extensions of a Lie group . In their Lie algebra formulation
they correspond to a curvature two-form (which is symmetric rather
than antisymmetric in the case of supersymmetry, see \cite{aa}).
The structure of the fractional supersymmetry group opens the possibility
of extending these concepts
to a (here) ternary algebra by introducing a
`curvature' three-form ({\it cf.} \cite{k}).

To exhibit the origin of (2.6), (2.8) and (2.11),
 we observe
 that in any context where such results arise, there is a natural
ordering of the numbers of non-zero grading that enter it. It will further
be seen  that  this ordering determines consistently (and
always according to the same pattern) the powers of $q$ that enter the
required multiplicative relations. In the case of group multiplication,
the above ordering (in {\it symbolic}
notation $\epsilon' > \epsilon > \theta)$ requires

$$
\epsilon' \epsilon = q^{-1} \epsilon \epsilon' \quad ,
\quad  \epsilon \theta = q^{-1} \theta \epsilon
\eqno{(2.12)}
$$

\noindent
used above. To these, we add the result
$\epsilon' \theta = q \theta \epsilon'$
(see below). For all three, one passes from the lexical order
to the opposite one by using relations that use the same power of $q$,
here $q^{-1}$, in the same places.
In the discussion of associativity, the ordering $\epsilon'' > \epsilon' >
\epsilon$
similarly implies the results (2.11) used above, and in addition $\epsilon''
\epsilon = q^{-1} \epsilon \epsilon''$.

Similarly, if we had
elect to write our fractional supersymmetry transformation as

$$
\theta' = \theta + \eta \quad , \quad t' = t + \tau + q^2 (\eta \theta^2 +
\eta^2 \theta) \quad ,
\eqno{(2.13)}
$$

\noindent
 the reality of $t'$ would now imply

$$
\eta \theta = q \theta \eta \quad ,
\eqno{(2.14)}
$$

\noindent
and, for the  ordering $\eta' > \eta > \theta$, the same rule would
govern matters but with the power $q$  as in (2.14), and in $\eta' \eta =
q \eta \eta'$. However, (2.13) is equivalent to

$$
\theta' = \theta + \eta \quad , \quad t' = t + \tau + q (\theta^2
\eta + \theta \eta^2) \quad ,
\eqno{(2.15)}
$$

\noindent
so that we prefer the ordering $\theta > \eta > \eta' $, and write

$$
\theta \eta =  q^{-1} \eta \theta \quad , \quad\eta \eta'= q^{-1} \eta' \eta
\quad , \quad
\theta \eta' = q^{-1} \eta' \theta \quad .
\eqno{(2.16)}
$$

\noindent
This is now in full conformity with the other examples
discussed Further, just as our discussion related to
$\epsilon' > \epsilon > \theta$
is appropriate to the case of left transformations, the passage involving
(2.15) and $\theta > \eta > \eta'$ is seen to be
similarly suited to the discussion of right translations. The results
(2.15) and (2.16) are indeed so employed in section six.

One consequence of results of the type (2.8) is in the form of $q$-deformed
binomial expansions. For example,

$$
(\epsilon' + \epsilon)^{m} = \sum_{t = 0}^{r} \left[
\begin{array}{c}
m\\
t
\end{array} \right] \epsilon'^{t} \epsilon^{m - t} \, .
\eqno{(2.17)}
$$

\noindent
The braced object here is the $q$-analogue of the ordinary binomial
coefficient,
in which ordinary factorials, {\it e.g.} $m!$, are replaced by

$$
[m] ! =  [m] [m - 1] \cdots [1] \quad  , \quad
[m] \equiv \frac{1 - q^m}{1 - q} =  1 + q + \cdots + q^{m - 1} \quad  .
\eqno{(2.18)}
$$

\noindent
It is easy to see and well-known that (2.17) indeed follows by use of (2.8).
Results such as (2.17) are employed in section three.

	We append our notation for $q$-deformed exponentials for use
in sections six and seven. We write

$$
\exp (q^k ; X) = \sum_{m = 0}^\infty \frac{1}{[m ; q^k]!} X^m \quad ,
\eqno{(2.19)}
$$

\noindent
for suitable $k$, where

$$
[m ; q^k]! = [m ; q^k] \cdots [2 ; q^k] [1] \quad , \quad
[m ; q^k] \equiv \frac{1 - q^{k m}}{1 - q^k} \quad .
\eqno{(2.20)}
$$

\noindent
In this notation $[m]$ in (2.18) is $[m ; q]$.

\section{The transformation formula for $G_n$}

We now extend the work done in the previous section on
$G_3$ to the ${\cal Z}_n$-graded case which employs
a single real Grassmann number $\theta$ and an associated parameter
$\epsilon$ with
the properties

$$
\theta = \bar{\theta} \quad , \quad \theta^n = 0 \quad ;
\quad \epsilon = \bar{\epsilon} \quad , \quad
\epsilon^n = 0 \quad ; \quad \epsilon \theta = q^{-1} \theta \epsilon \quad .
\eqno{(3.1)}
$$

\noindent
It is understood that no power of $\theta$ or $\epsilon$ lower than the
$n$-th can
vanish.

We retain the general structure (2.4) for $G_n$ but seek, for the
cocycle $\xi$, a formula of the type

$$
\xi (\epsilon', \epsilon) = \sum_{r = 1}^{n - 1} c_r \epsilon'\,^r
\epsilon^{n-r}
q^{\omega(r)} \quad ,
$$

$$
q =\exp  (2 \pi i /n)\quad , \quad n = 2,3,4 ... \quad ,
\eqno{(3.2)}
$$

\noindent
so that $q^n  = 1 $
replaces $q^3 = 1$
in previous work. Also, the exponent of $q$ shown in (3.2)
namely

$$
\omega (r) = \frac{1}{2} r (n - r) \quad,
\eqno{(3.3)}
$$

\noindent
ensures using (2.8) that each term of (3.2) is real if $c_r$ is real.
We set $c_1=1$, and rewrite (3.2) as
$$\xi(\epsilon',\epsilon)=\sum_{r=1}^{n-1} d_r \epsilon'\, ^r \epsilon^{n-r}
\quad .
\eqno{(3.4)}
$$
We must determine the numbers $d_r$ in such a way that (2.10) is satisfied,
so that when $\xi$ is given by (3.2) and (3.3), eq. (2.4) has
 the required $G_n$ group multiplication
properties. First we note that the terms on the two sides of (2.10) that are
independent of $\epsilon$ agree. Then, with the aid of results like (2.17), we
can show that consideration of the terms of (2.10) linear in $\epsilon$
allow us to determine all the $d_r$ as multiples of $d_1$. Explicitly we find
$d_r=d_{n-r}$ and
$$
\frac{d_r}{d_1}=\frac{[n-1]!}{[r]![n-r]!} \quad , \quad r=1,2, \dots ,n-1.
\eqno{(3.5)}
$$

\noindent
Now that (3.4) is fully determined by (3.5) we must prove that (2.10) is
identically satisfied. Thus, we use (3.4), (3.5) and (2.17) to obtain
$$
\xi(\epsilon'',\epsilon'+\epsilon)=
\sum_{r=1}^{n-1} \sum_{s=0}^{n-r} \frac {\epsilon''\, ^r \epsilon'\, ^s
\epsilon^{n-r-s} d_1 [n-1]!}{[r]! [s]! [n-r-s]!} \quad .
\eqno{(3.6)}
$$
\noindent
We now observe that $\xi(\epsilon',\epsilon)$ differs only slightly
from what will provide the $r$=0 of the r.h.s. of (3.6). In fact, we
can write
$$
\xi(\epsilon'',\epsilon'+\epsilon)+\xi(\epsilon',\epsilon)=
\sum_{r=0}^n \sum_{s=0}^{n-r} \frac {\epsilon''\, ^r \epsilon'\, ^s
\epsilon^{n-r-s} d_1 [n-1]!}{[r]! [s]! [n-r-s]!}
-d_1(\epsilon''^{n}+\epsilon'^{n}+\epsilon^{n})/[n] \quad.
\eqno{(3.7)}
$$
\noindent
Here, in order to make a tractable double sum we have added and subtracted
certain ill-defined terms. The procedure is necessary to expedite the
key step of our proof. In this, we reverse the order of summations in (3.7)
obtaining
$$\sum_{r=0}^n \sum_{s=0}^{r}=\sum_{s=0}^n \sum_{r=s}^n=
\sum_{s=0}^n \sum_{u=0}^{n-s} \quad , $$
\noindent
where a shift in the variable of summation $r$ to $u=r-s$ has also been made.
The result so obtained for the left side of (2.10) can now be shown to agree
exactly with the analogue of (3.7) obtained by direct calculation
of the r.h.s. of (2.19), completing
the required demonstration.

The remaining ingredients of the group multiplication laws for $G_n$ are
attended to immediately. Indeed, an additional calculation to prove
associativity is not needed, since the cocycle property guarantees it.

\section{Derivatives with respect to $\theta$}

We want to move from the description of the group properties of
the fractional supersymmetry transformation towards the construction of
actions and dynamical systems that possess invariance properties relative
to them. This requires a geometrical understanding of the derivatives
$\partial / \partial \theta$,
and of objects in the theory of covariant derivative type. Let us
go back to  $G_3$, again as a good example,
aiming in particular to expose and treat the conceptual difficulties
that occur in discussing reality properties of $\partial/\partial \theta$.
It is sufficient for the purposes of this section, although not of course
for the eventual construction  of Lagrangian theories, to work with scalar,
{\it i.e.}  grade zero real superfields $f$,
whose expansion in powers of $\theta$
involves three real terms (see {\it  e.g.} \cite{adl,dur,de,cm})

$$
f = x + q \alpha \theta + q \beta \theta^2  \; \;  = \; \; \bar{f} =
x + q^2 \theta \alpha + q^2 \theta^2 \beta \quad  ,
\eqno{(4.1)}
$$

\noindent
in which $x$ is a grade zero (bosonic) variable, and the variables $\alpha$
and $\beta$ are of grades two and one. The reality of $f$ expressed by
 (4.1)  implies the properties

$$
\theta \beta = q \beta \theta \quad ,
\quad \theta \alpha = q^{-1} \alpha \theta \quad .
\eqno{(4.2)}
$$

\noindent
Comparing (4.1) with (2.5) now seen to be of scalar superfield nature,
we see that $\beta$, of grade one, is related like $\epsilon$ to
$\theta$, so that $\beta > \theta$, and $\alpha$, of grade two, is likewise
related to $\epsilon^2$, so that $\alpha < \theta$. The latter implies
that we should adopt the rule $\beta \alpha = q^{-1} \alpha \beta$ , although
in this section no call for any such result is made.

To prepare the ground
for our discussion of derivatives in the ${\cal Z}_3$ case, we recall
 briefly  the case of basic supersymmetry and ${\cal Z}_2$-grading (eq. (11))
for which a  the real
scalar field  has the expansion

$$
f = x + i \theta \phi = x - i \phi \theta \quad .
\eqno{(4.3)}
$$

\noindent
It is normal to use the left spinorial derivative so that

$$
\frac{\partial f}{\partial \theta} = \partial f \equiv
\partial_L  f = i \phi \quad ,
\eqno{(4.4)}
$$

\noindent
and to employ $\frac{\partial \theta}{\partial \theta} = 1$ and

$$
\theta \partial + \partial \theta = 1
\eqno{(4.5)}
$$

\noindent
to do routine manipulations. Since (4.4) is not real for real $f$
there is no case for viewing $\partial$ as a real entity. However
one did not consider using such an idea as a  guide  towards (4.5).
Eq.  (4.5) is valid  because it holds  applied to an arbitrary
superfield $f$. In fact, the right spinorial derivative
$\partial_R$
can be consistently viewed as a conjugate to $\partial_L$ via

$$
\overline{\partial_L f} \equiv
\frac{\overline{\partial f}}{\partial \theta} \equiv
f \frac{\stackrel{\leftarrow}{\partial}}{\partial \theta} = \partial_R f
\quad ,
\eqno{(4.6)}
$$

\noindent
which agrees trivially with

$$
\partial_L f = i \phi \quad , \quad \partial_R f = - i \phi \quad .
$$

\noindent
Similarly, by application to arbitrary $f$ it follows that the
conjugate of (4.5),

$$
\partial_R \theta + \theta \partial_R = 1 \quad ,
$$

\noindent
makes good sense.

Returning to the ${\cal Z}_3$ case, we see that to compute
$\frac{\partial f}{\partial \theta} \equiv \partial f \equiv \partial_L f$
 and
$\partial_R f$, we need the ${\cal Z}_3$-analogue of (4.5) to treat the
$\theta^2$ terms of (4.1). We begin by postulating

$$
\frac{\partial \theta}{\partial \theta} = 1
\eqno{(4.7)}
$$

\noindent
and a result of the type

$$
\partial \theta = a \theta \partial + b \quad ,
\eqno{(4.8)}
$$

\noindent
in which $a, b \in  C$. Eq. (4.7) is certainly natural. We discuss
whether it can or needs to be modified (it doesn't) below. When (4.8)
is applied to 1, then (4.7) implies $b =1$.
 Applied to $\theta$, eq.  (4.8) yields

$$
(\partial \theta^2) = ( 1 + a) \theta \quad ,
\eqno{(4.9)}
$$

\noindent
Then, using (4.1), we get

$$
\frac{\partial f}{\partial \theta} = \partial_L f = q^2 \alpha
+ q^2 (1 + a) \theta \beta \quad  ,
\eqno{(4.10)}
$$

\noindent
which is  not real  for real $f$. To complete the
specification of (4.8), we stipulate that it must be  a true
 when applied to an arbitrary scalar superfield. It is easy to see
that it does so if

$$
1 + a + a^2 = 0 \quad .
\eqno{(4.11)}
$$

\noindent
Thus we find two solutions for $a$; the two corresponding candidates for
the derivative with respect $\theta$ are both used in the literature and, as
we see below, essential. If $a = q$, we shall write $\partial$ for the
derivative that obeys

$$
\frac{\partial \theta}{\partial \theta} = 1 \quad , \quad \partial \theta
= q \theta \partial + 1 \quad , \quad [\partial , \theta]_q = 1 \quad .
\eqno{(4.12)}
$$

\noindent
If $a = q^{-1}$, we write $\delta$, and

$$
\frac{\delta \theta}{\delta \theta } = 1 \quad , \quad
 \delta \theta = q^{-1} \theta
\delta + 1 \quad ,  \quad [\delta , \theta]_{q^{-1}} = 1 \quad .
\eqno{(4.13)}
$$

\noindent
Also $\partial \delta = q^{-1} \delta \partial $ (or $[\partial,
\delta]_{q^{-1}} = 0$). Both derivatives
hereby introduced are acting from the left. Neither has any natural reality
properties  that can be uncovered without reference to their partner right
derivatives. The above is sufficient for
our own intended  applications. However, variations in the literature
exist, and are often associated with implicit assumptions hinting at
reality properties of $\partial$. If one uses (4.8)
with or without (4.7) and without reference to the
requirement that, applied to an arbitrary superfield, it holds good,
one might try to complete specification
of (4.8) by demanding that its correctness ensures the correctness of
its adjoint. However,  if one assumes $\partial$ is in
some sense real (which we do not believe to be a tenable view) then
(4.8) implies successively

$$
\begin{array}{ll}
\theta \partial & = \bar{a} \partial \theta + \bar{b}\quad, \\
a \theta \partial & = a \bar{a} \partial \theta + a \bar{b}\quad ,
\hbox{ calling for } \quad a \bar{a} = 1\quad, \\
\partial \theta & =  a \theta \partial - a \bar{b} \quad ,
\end{array}
$$

\noindent
reproducing (4.8) when $b = - a \bar{b}$. The choice $b = 1$,
the natural choice, implies $a = -1$, and we are
forced back to the ${\cal Z}_2$-supersymmetry result
as the only non-trivial possibility.
If one tries a choice like $a =
q$ , then $b = i q^{1/2} r\, , \,  r \in  R$, so that

$$
\partial \theta = q \theta \partial + i r q^{1/2} \quad .
$$

\noindent
Application of this result to an arbitrary real scalar $f$ fails
to give an identity. So also does {\it any} attempt to view $c \partial$
as a conjugate to $\partial$, for $c \in C$.

In fact, it is sensible to view $\partial_R$ as the conjugate of $
\partial$. Since doing so is independent of whether one is looking
at $\partial$ or $\delta$, it is sufficient to give details for the
former. Thus we shall employ here (4.12) and (4.10). We take
$\partial_R \theta = ( \theta \frac{\stackrel{\leftarrow}{\partial}}
{\partial \theta}) = 1$
and, from (4.9), by conjugation,
deduce

$$
(\partial_R \theta^2 ) \equiv (\theta^2 \frac{\stackrel{\leftarrow}{\partial}}
{\partial \theta})
= (q^2 + 1) \theta \quad .
$$

\noindent
Then, from (4.1), we obtain

$$
\partial_R f = (f \frac{\stackrel{\leftarrow}{\partial}}{\partial \theta})
=  q
\alpha + q \beta (1 + q^2) \theta = \overline{\partial_L f}
\eqno{(4.14)}
$$

\noindent
where (4.2) and (4.10) for $\partial f \equiv \partial_L f$ have been
used. Similarly, a consistent picture for $\delta , \delta_R$ emerges.
So, in summary, if on rare occasions one needs a conjugate for $\partial$,
one may not use any multiple of $\partial$, although
$\partial_R$ serves  perfectly well. Neither
$\delta $ nor $\delta_R$ are satisfactory
candidates
for the  r\^ole of the conjugate of $\partial$. We note in passing
that the fact that a variable and
the derivative with respect to it cannot be made {\it simultaneously}
real (or hermitian) is a known feature of non-commutative
geometry and has been discussed in completely different contexts (see,
{\it e.g.} \cite{OSWZ}).

We note that the result

$$
q \partial_L \partial_R f = \partial_R \partial_L f \quad  ,
$$

\noindent
treated with care, also makes sense, but forbear from appending any remark
about ordering.

We return finally to (4.7). It is not obviously wrong to let
$(\partial \theta) = c , c \in C$, but (4.7) clearly remains
the natural choice. With $\partial_R$, rather than any multiple of $\partial$,
seen as the true conjugate of $\partial$, we are not aware of any compelling
reason for using $c \neq 1$.

\section{ Covariant derivative objects}

The derivatives $\partial$  and $\delta$ discussed  feature in
the literature on fractional supersymmetry in the definition (see
{\it e.g.} \cite{adl,dur,cm}) of the important quantities

$$
Q = \partial_\theta + q \theta^2 \partial_t \quad  ,
\eqno{(5.1)}
$$

$$
D= \delta_\theta + q^2 \theta^2 \partial_t \quad ;
\eqno{(5.2)}
$$

\noindent
 $Q$ produces the first order generalised supersymmetry
transformation. We then write

$$
\delta_{(\epsilon)} f = \epsilon Q f \quad .
\eqno{(5.3)}
$$

\noindent
Eq. (5.3) implies the superfield component transformations

$$
\begin{array}{ll}
\delta_{(\epsilon)} x &= q^2 \epsilon \alpha \quad , \\
\delta_{(\epsilon)} \alpha &= -q \epsilon \beta \quad , \\
\delta_{(\epsilon)} \beta &= \epsilon \dot{x} \quad .
\end{array}
\eqno{(5.4)}
$$

\noindent
We note that the $`\theta^2$ component' of $f$ changes by a total time
derivative.
Proceeding from this remark towards the construction of actions, we realise
 that
$D$ has been defined in (5.2) and in relation to (5.1) in such a way that
$D f$
has the same transformation law as $f$. It follows that the same philosophy as
worked for supersymmetry will enable us to construct actions with the correct
invariance properties under generalised supersymmetry transformations. We just
take the `$\theta^2$ component'
of a suitable product, of the correct dimensions,
of superfields such as $f \, , \dot{f} \, , Df$ etc. For example ({\it cf.}
\cite{dur,de,cm})

$$
S = \int \,  dt \,  \frac{1}{2} \, i \,  f \, D \, f \, |_{\theta^2} =
\int \, dt \, L \quad ,
\eqno{(5.5)}
$$

$$
L = \frac{1}{2} \dot{x} \dot{x} + \frac{1}{2} q^2 \dot{\beta} \alpha
- \frac{1}{2} q \dot{\alpha} \beta \quad .
\eqno{(5.6)}
$$

\noindent
Exposition of the canonical formalism that stems from (5.6) is neither problem
free in quantum mechanics, nor in existence at all at the present time to our
knowledge in classical mechanics. We may expect,
as is the case in ${\cal Z}_2$-supersymmetry where symmetric Poisson
brackets are associated with anticommutators, that both formalisms,
classical (fractional pseudomechanics) and quantum, are closely related.
We intend to present a discussion of these questions elsewhere.

The important r\^oles of $Q$ and $D$ having been put into evidence,
we note that $D f$ will transform like $f$ provided that

$$
\delta_{(\epsilon)}    D  \;  = \;  D  \delta_{(\epsilon)} \quad ,
\eqno{(5.7)}
$$

\noindent
or

$$
\epsilon \, Q \, D \; = \; D \, \epsilon \,  Q  \quad .
\eqno{(5.8)}
$$

\noindent
Since (2.6) implies $\theta^2 \, \epsilon = q^2 \, \epsilon \,
\theta^2 \, , $ and hence

$$
q \, D \, \epsilon = \epsilon \, D \, ,
$$

\noindent
we deduce that (5.7) requires

$$
D \, Q = q \, Q \, D \quad , \quad [D, Q]_q = 0 \quad .
\eqno{(5.9)}
$$

\noindent
The consistency of (5.8) as an operator identity demands that, when
applied to an arbitrary $f$, it gives a superfield identity. The choice (5.2)
shows this to be satisfied.

To conclude, we note the further well-known results ({\it cf.}
\cite{adl,dur,cm})

$$
D^3 = -\partial_t \quad,\quad Q^3 = -\partial_t \quad ,
\eqno{(5.9)}
$$

\noindent
which are most easily seen as identities by applying them to arbitrary $f$.

\section{Left and Right Transformations}

What governed the choices (5.1), (5.2)? In the case of (5.1) the application of
$\epsilon Q$ to $(t,\theta)$ does reproduce (2.4) to {\it first} order
in $\epsilon$. This however does not allow $\partial$ to be preferred
to $\delta$ in (5.1), nor conversely. Once (5.1) has been chosen, as
seems sensible enough, it is quite easy  to find a derivative in the
form (5.2) that satisfies (5.8). However, this choice has a deep
geometrical  interpretation. In fact,
we now show that $Q$ and $D$  can be regarded as the generators of the left
 and
right actions of the group $G_3$ on the physical `manifold' $M$ of (2.9),
 and
that (5.8) expresses the fact that  left and right actions commute. This
 geometrical picture is a nontrivial generalisation of one that
applies
to ordinary supersymmetry where, of course,
both actions are linear.

Let us  denote the parameters of the left and right transformations of
$G_3$ as $\epsilon$ and $\eta$. Thus
$\epsilon > \theta > \eta$, as
discussed in section two, and hence

$$
\epsilon \theta = q^{-1} \theta \epsilon \quad, \quad
\theta \eta = q^{-1} \eta \theta \quad .
\eqno{(6.1)}
$$

\noindent
We define the left and right actions $L_{(\epsilon)}$ and
$R_{(\eta)}$ by

$$
L_{(\epsilon)} : \theta \mapsto \theta ' = \epsilon + \theta \quad ,
\quad t \mapsto
 t' = t + \tau + q (\epsilon \theta^2 + \epsilon^2 \theta) \quad ,
\eqno{(6.2)}
$$

$$
R_{(\eta)} : \theta \mapsto \theta' = \theta + \eta \quad , \quad
 t \mapsto t' =
t + \tau + q (\theta^2 \eta + \theta \eta^2) \quad ,
\eqno{(6.3)}
$$

\noindent
which agree with (2.4) and (2.15). It is a non-trivial result
is that these transformations may written as exponentials of the
generators $Q$ and $D$ respectively

$$
L{(\epsilon)}\; t = \exp\;(q^{-1} \; ; \epsilon Q) \,  t\quad,
\eqno{(6.4)}
$$

$$
R_{(\eta)}\; t = \exp \;(q \; ; \eta  D) \, t\quad,
\eqno{(6.5)}
$$

\noindent
where we have used the notation of (2.19).
The proof, which due to the ordering necessarily
involves distinct deformed exponentials,
is given below.  The
commutativity of the two actions implies

$$
[ \epsilon Q \; , \; \eta D ] = 0 \quad .
\eqno{(6.6)}
$$

\noindent
This requires the consequences

$$
D \epsilon = q^{-1} \epsilon D \quad , \quad \eta Q = q^{-1} Q \eta
\eqno{(6.7)}
$$

\noindent
of (6.1), and a hitherto unused relation
$$
\epsilon \eta = q^{-1} \eta \epsilon \quad .
\eqno{(6.8)}
$$

\noindent
Then (6.6) is seen to imply (5.8).

To prove (6.4), we use (2.19) in the form

$$
 \exp (q^{-1}; \epsilon Q)
= 1 + \epsilon Q + \epsilon Q \epsilon Q / [2 \; ; \; q^{-1}] \quad .
\eqno{(6.9)}
$$

\noindent
A simple computation using $\epsilon \theta = q^{-1} \theta \epsilon$ and
$(\partial
\theta^2) = [ 2 ; q ] \theta$ gives us (6.4); the proof depends crucially
on the former and on the occurrence of $\partial$ rather than $\delta$
in the definition (5.1) of $Q$.

We prove (6.5) in the same way, noting again how critically the success
of the proof depends on the actual arrangement of details involving
$\theta , \delta , \eta$ and $\exp \;(q\; ; \;\eta D).$

\section{Superfields, derivatives and
$q$-exponentiation for $G_n$}

We have given already in section three, the definition of the transformation
of the group $G_n$ when

$$ \theta = \bar{\theta} \quad , \quad \theta^n = 0 \quad ; \quad \epsilon
= \bar{\epsilon} \; , \epsilon^n
= 0 \quad ; \quad \theta \epsilon = q \epsilon \theta \quad .
\eqno{(7.1)}
$$

\noindent
In general we expect most features of the ${\cal Z}_3$ - graded theory that
 are
discussed above allow fairly direct extension to the ${\cal Z}_n$ theory. We
will indicate some of these briefly in this section, without examining in
much
detail how the general case may yield a theory significantly richer in content.

In place of (4.1), we have the expansion of the real scalar superfield

$$
f = x + \sum_{r = 1}^{n - 1} q^{\omega (r)} \psi_r \theta^{n - r} =
\bar{f}
= x + \sum_{r = 1}^{n - 1} q^{- \omega (r)} \theta^{n - r} \psi_{r}
\quad ,
\eqno{(7.2)}
$$

\noindent
where the power $\omega (r)$ of $q = \exp (2 \pi i /n)$ is
chosen to make all terms in $f$ all real; it is given by (3.3).
Moreover, in place of (4.2), we now have
$$
\theta \psi_r = q^r \psi_r \theta \; , \; r = 1 , \dots,   n - r
\quad ;
\eqno{(7.3)}
$$
\noindent
in the ${\cal Z}_3$ case, $\alpha$ and $\beta$ of (4.1) would be written as
$\psi_2 $ and $\psi_1$ to conform with (7.2).

	The discussion of derivatives, via (7.1) and

$$
\partial \theta = a \theta \partial + 1 \quad ,
\eqno{(7.4)}
$$

\noindent
yields more possibilities, for $a$ must now obey

$$
1 + a + a^2 + ... \; + a^{n-1} = 0 \quad .
\eqno{(7.5)}
$$

\noindent
We thus write $\partial_r$ for the derivative which obeys

$$
\partial_r \; \theta = q^r \; \theta \;  \partial_r + 1 \quad  ,
\quad  r = 1, 2 ... \; n - 1 \quad .
\eqno{(7.6)}
$$

\noindent
Our previous $\partial$ and $\delta$ correspond to $\partial_{1}$
and $\partial_{n -1}$.
 We will continue to use the former notation because we
do not describe any context that involves crucial use of $\partial_r$ for
$r \neq 1$ or $r\neq n - 1$. We note, in particular,
the direct consequences of (7.6)

$$
(\partial \, \theta^s) = [s] \theta^{s-1} \quad , \quad
(\delta \, \theta^s) = q^{1-s} [s] \theta^{s-1} \quad , \quad
 s = 1, 2, ... \;, n - 1 \quad .
\eqno{(7.7)}
$$

\noindent
Here $\delta$ directly involves $(1 - q^{-s}) / (1 - q^{-1})$, which we
have expressed in terms of $[s]$, defined by (2.18).

The definitions (5.1) and (5.2) of $Q , D$ in section five are now
modified to read

$$
Q = \partial_1 + q^{\omega(1)} \theta^{n-1} \partial_t
\equiv \partial + q^{\omega(1)} \theta^{n-1} \partial_t \quad ,
\eqno{(7.8)}
$$

$$
D = \partial_{n-1} + q^{-\omega(1)} \theta^{n-1} \partial_t
\equiv \delta + q^{-\omega(1)} \theta^{n-1} \partial_t \quad ,
\eqno{(7.9)}
$$

\noindent
where $\omega(1)$ is given by (3.3).
We note that $Q$, so defined, does generate correctly the first order term
of the $G_n$-transformation  for  a parameter
$\epsilon $ related to $\theta$ via (7.1). Also

$$
\delta_{(\epsilon)} \psi_{1} = \epsilon \dot{x}
\eqno{(7.10)}
$$

\noindent
indicates that we can still follow the usual way of obtaining invariant actions
from the $\theta^{n-1}$ components of suitable superfields. Further (5.8)
again holds.
 But in place of (5.9), we use $n$-th powers: the ${\cal Z}_n$ theory
is of fractional supersymmetry with fractions $1/n$, of course.

Finally, it is to be expected that the exponentiation results of section
six carry over into the general theory. We rewrite then in general notation

$$
L_{(\epsilon)} t \mapsto t'
= \exp ( q^{-1}  ; \epsilon Q) \,t \quad  ,
\eqno{(7.11)}
$$

$$
R_{(\eta)} t \mapsto t' = \exp  ( q ; \eta D)\,t  \quad  ,
\eqno{(7.12)}
$$

\noindent
where

$$
\theta \epsilon = q \epsilon \theta \quad ,
\quad  \eta \theta = q \theta \eta \quad , \quad
\eta \epsilon = q \epsilon \eta \quad .
\eqno{(7.13)}
$$

\noindent
To prove the extension (7.11) of (6.4) to the ${\cal Z}_n$-graded
theory we need to recover

$$
L_{(\epsilon)} : t \rightarrow t' = t + \sum_{r = 1}^{n - 1} d_r
\epsilon^r \theta^{n-r} \quad  ,
\eqno{(7.14)}
$$

\noindent
where $d_r$ is given by (3.5), as the expansion  of (7.11)

$$
L_{(\epsilon)} t = t +  \sum_{r = 1}^n \frac{1}{[r ; q^{-1}]!}
(\epsilon Q)^r t \quad  ,
\eqno{(7.15)}
$$

\noindent
where $Q$ is given by (7.8). The first order term,
which comes from the  action of $\epsilon  Q$ on
$t$ is clearly correct:

$$
d_1 \epsilon\theta^{n - 1} = (\epsilon Q) t \quad ,
\eqno{(7.16)}
$$

\noindent
since $d_1 = c_1 q^{\omega (1)} = q^{\omega (1)}$.

 This is first key element
of a proof, by induction, that the individual terms of (7.14) and (7.15)
coincide. We therefore assume this for $r = 1, 2, ..., k$ and seek, on the
basis of that assumption, to prove it for $r = k + 1$. This requires
us to show that

$$
d_{k + 1} \epsilon^{k + 1} \theta^{n - k -1} =
\frac{1}{[k + 1 ; q^{-1}]} (\epsilon Q) d_k \epsilon^k \theta^{n - k}
\quad .
\eqno{(7.17)}
$$

\noindent
Only the term $\partial_1 \equiv \partial$ of $Q$ contributes. The
formula $\partial \epsilon^k = q^{-k} \epsilon^k \partial$
then prepares for the use of (7.7), and we can see that (7.17) is an
 equality  provided that

$$
\frac{1}{[k + 1 ;q]} = \frac{q^{-k }}{[k + 1 ; q^{-1}]} \quad .
\eqno{(7.18)}
$$

\noindent
It is easy to show that (7.18) is true, and the proof is complete.

Proof of (7.12) proceeds similarly. We remark that the exponentials
in (7.11) and (7.12) are necessarily different because of the use
of the different derivatives $\partial$ and $\delta$ in the definitions
(7.8) and (7.9) of $Q$ and $D$.

\vspace{2cm}
\noindent
{\bf Acknowledgements}. This research has been supported in part by a grant
from the CICYT, Spain.
A.J. Macfarlane  also wishes to thank the Department of Theoretical Physics
of Valencia University for its hospitality.

\end{document}